\documentclass[aps,prd,preprint,showpacs,showkeys,superscriptaddress,nofootinbib]{revtex4-1}

\usepackage{booktabs}
\usepackage{graphicx}
\usepackage{xcolor}
\usepackage{amsmath}
\usepackage{hyperref}
\usepackage{bm}
\usepackage{mathtools}
\usepackage{amssymb}
\usepackage{lipsum}
\usepackage{epstopdf}
\usepackage{slashed}
\usepackage{multirow}
\usepackage{appendix}

\AtBeginDocument{
\heavyrulewidth=.08em
\lightrulewidth=.05em
\cmidrulewidth=.03em
\belowrulesep=.65ex
\belowbottomsep=0pt
\aboverulesep=.4ex
\abovetopsep=0pt
\cmidrulesep=\doublerulesep
\cmidrulekern=.5em
\defaultaddspace=.5em
}

\allowdisplaybreaks

\begin{document}
\def\qq{\langle \bar q q \rangle}
\def\uu{\langle \bar u u \rangle}
\def\dd{\langle \bar d d \rangle}
\def\sp{\langle \bar s s \rangle}
\def\GG{\langle g_s^2 GG \rangle}
\def\Tr{\mbox{Tr}}

\def\lrar{\leftrightarrow}  

\def\ds{\displaystyle}
\def\beq{\begin{equation}}
\def\eeq{\end{equation}}
\def\bea{\begin{eqnarray}}
\def\eea{\end{eqnarray}}
\def\beeq{\begin{eqnarray}}
\def\eeeq{\end{eqnarray}}
\def\ve{\vert}
\def\vel{\left|}
\def\ver{\right|}
\def\nnb{\nonumber}
\def\ga{\left(}
\def\dr{\right)}
\def\aga{\left\{}
\def\adr{\right\}}
\def\lla{\left<}
\def\rra{\right>}
\def\rar{\rightarrow}
\def\nnb{\nonumber}
\def\la{\langle}
\def\ra{\rangle}
\def\ba{\begin{array}}
\def\ea{\end{array}}
\def\tr{\mbox{Tr}}
\def\ssp{{\Sigma^{*+}}}
\def\sso{{\Sigma^{*0}}}
\def\ssm{{\Sigma^{*-}}}
\def\xis0{{\Xi^{*0}}}
\def\xism{{\Xi^{*-}}}

\def\qs{\la \bar s s \ra}
\def\qu{\la \bar u u \ra}
\def\qd{\la \bar d d \ra}

\def\gGgG{\la g^2 G^2 \ra}
\def\q{\gamma_5 \not\!q}
\def\x{\gamma_5 \not\!x}
\def\g5{\gamma_5}
\def\sb{S_Q^{cf}}
\def\sd{S_d^{be}}
\def\su{S_u^{ad}}
\def\sbp{{S}_Q^{'cf}}
\def\sdp{{S}_d^{'be}}
\def\sup{{S}_u^{'ad}}
\def\ssp{{S}_s^{'??}}

\def\sig{\sigma_{\mu \nu} \gamma_5 p^\mu q^\nu}
\def\fo{f_0(\frac{s_0}{M^2})}
\def\ffi{f_1(\frac{s_0}{M^2})}
\def\fii{f_2(\frac{s_0}{M^2})}
\def\O{{\cal O}}
\def\sl{{\Sigma^0 \Lambda}}
\def\es{\!\!\! &~=~& \!\!\!}
\def\ap{\!\!\! &\approx& \!\!\!}
\def\ar{&+& \!\!\!}
\def\ek{&-& \!\!\!}
\def\kek{\!\!\!&-& \!\!\!}
\def\cp{&\times& \!\!\!}
\def\se{\!\!\! &\simeq& \!\!\!}
\def\eqv{&\equiv& \!\!\!}
\def\kpm{&\pm& \!\!\!}
\def\kmp{&\mp& \!\!\!}
\def\mcdot{\!\cdot\!}
\def\erar{&\rightarrow&}


\title{$\gamma^* N \rightarrow \Delta^{+}(1600)$ transition form factors in light-cone sum rules}

\author{T.~M.~Aliev}
\email{taliev@metu.edu.tr}
\affiliation{Department of Physics, Middle East Technical University, Ankara, 06800, Turkey}

\author{T.~Barakat}
\email{tbarakat@ksu.edu.sa}
\affiliation{Department of Physics, King Saud University, Riyadh, 11451, Saudi Arabia}

\author{S.~Bilmis}
\email{sbilmis@metu.edu.tr}
\affiliation{Department of Physics, Middle East Technical University, Ankara, 06800, Turkey}

\author{M.~Savci}
\email{savci@metu.edu.tr}
\affiliation{Department of Physics, Middle East Technical University, Ankara, 06800, Turkey}

\date{\today}

\begin{abstract}

  The form factors of $\gamma^* N \rightarrow \Delta(1600)$ transition is calculated within the light-cone sum rules assuming that $\Delta^+(1600)$ is the first radial excitation of $\Delta(1232)$. The $Q^2$ dependence of the magnetic dipole $\tilde{G}_M(Q^2)$, electric quadrupole $\tilde{G}_E(Q^2)$, and Coulomb quadrupole $\tilde{G}_c(Q^2)$ form factors are investigated. Moreover,  the $Q^2$ dependence of the ratios $R_{EM} = -\frac{\tilde{G}_E(Q^2)}{\tilde{G}_M{(Q^2)}}$ and $R_{SM} = - \frac{1}{4 m_{\Delta(1600)}^2} \sqrt{4 m_{\Delta(1600)}^2 Q^2 + (m_{\Delta(1600)}^2 - Q^2 -m_N^2)^2} \frac{\tilde{G}_c(Q^2)}{\tilde{G}_M(Q^2)}$ are studied. Finally, our predictions on $\tilde{G}_M(Q^2)$, $\tilde{G}_E(Q^2)$, and $\tilde{G}_C(Q^2)$ are compared with the results of other theoretical approaches.
\end{abstract}

\maketitle

\section{Introduction}
Advance technologies in accelerators enabled to search of high energy regions as well as improving the precision of the experiments by collecting data with high luminosity. This reveals new possibilities to study the electromagnetic structures of the baryon resonances above the ground state region. The facilities like CLAS (Jefferson Lab), BATES (MIT), MAMI (Mainz), Spring-8 (Japan), and ESSA (Bonn) have the potential to measure the electromagnetic structures of baryons around their first excitations. These experimental possibilities stimulated theoretical studies for deeply understanding the properties of the baryon resonances. $\Delta(1600)$ baryon, which is the first excitation of $\Delta(1200)$ one, may be one of the resonances which deserves special attention. Theoretically, this resonance has not been studied comprehensively yet.

The electroproduction of $\Delta(1600)$ is studied within the quark model~\cite{Capstick:1994ne}, and the effects of $\Delta(1600)$ in baryon-meson reactions is studied in~\cite{Burkert:2019opk,Golli:2007sa}. However, the existing data~\cite{Trivedi:2018rgo,Burkert:2019opk} can be used for a more detailed analysis of this resonance. The $\gamma^* p \rightarrow \Delta^+(1232),~(\Delta^+(1600))$ transitions are computed using a diquark-quark picture and a covariant spectator constituent quark model in~\cite{Lu:2019bjs} and \cite{Ramalho:2010cw}, respectively.

The form factors of the $\gamma^* N \rightarrow \Delta(1232)$ and $\gamma^*~\text{octet} \rightarrow \text{decuplet}$ baryon transitions within the same framework was studied in~\cite{Braun:2005be} and \cite{Aliev:2013jta}, respectively. In present work, we study the transition form factors for the electroproduction of the $\Delta(1600)$ resonance within light-cone sum sum rules method.

The article is organized as follows. In section~\ref{sec:2}, the sum rules for the transition form-factors of $\gamma^* N \rightarrow \Delta(1600)$ within the light-cone sum rules (LCSR) is derived. The numerical analysis of the obtained LCSRs is carried out in section~\ref{sec:3}. This section also contains discussion and summary.

\section{Derivation of LCSR for $\gamma^* N  \rightarrow \Delta(1600)$ transition form factors}
\label{sec:2}
The transition $\gamma^* N \rightarrow \Delta (1200)$ and $\Delta(1600)$ is described by the matrix element of the electromagnetic current $j_\mu = e_u \bar{u} \gamma_\mu u + e_d \bar{d} \gamma_\mu d$ between the nucleon, ground, and first radial excitation of $\Delta$ baryon $\langle \Delta_i(p^\prime) | j_\nu(0) | N(p) \rangle$.

By using the  Lorentz invariance and current conservation, this matrix element is determined in terms of the following form factors ~\cite{Devenish:1975jd}:

\begin{equation}
  \label{eq:1}
  \begin{split}
    \langle \Delta_i(p^\prime | j_\mu^{el})| N(p) \rangle =&~\bar{u}_\alpha(p^\prime) \Bigl\{ G_1^{(i)}(Q^2) (- q_\alpha \gamma_\mu + \slashed{q} g_{\alpha \mu}) + G_2^{(i)}(Q^2)(-q_\alpha \mathcal{P}_\mu + (q \mathcal{P}) g_{\alpha \mu}) \\
    &+
  G_3^{(i)} (q_\alpha q_\mu - q^2 g_{\alpha \mu})  \Bigr\} \gamma_5 u(p)
\end{split}
\end{equation}
where $i = 0$, and $i=1$ corresponds to the ground and first radial excitation of $\Delta$ baryon, $G_1^i$, $G_2^i$, and $G_3^i$ are the corresponding form-factors, and $\mathcal{P}_\alpha = \frac{1}{2}(p + p^\prime)_\alpha = \frac{1}{2}(2p^\prime +q)_\alpha$. However, the multipole form factors are more useful than the form factors $G_1$, $G_2$, and $G_3$ for the experimental point of view.
The relations among the form factors $G_1(Q^2)$, $G_2(Q^2)$, and $G_3(Q^2)$ and multipole form factors (magnetic dipole $G_M$, electric quadrupole $G_E$, and Coulomb quadrupole $G_c$) form factors are given in~\cite{Jones:1972ky}:
\begin{equation}
  \label{eq:2}
  \begin{split}
    G_M^{(i)} (Q^2) &= \frac{m_N}{3(m_N + m_{\Delta_i})} \big[ \big( (3m_{\Delta_i} + m_N)(m_{\Delta_i} + m_N) + Q^2 \big) \frac{G_1^{(i)}(Q^2)}{m_{\Delta_i}} \\
    &+ (m_{\Delta_i}^2 - m_N^2) G_2^{(i)}(Q^2) - 2Q^2 G_3^{(i)}(Q^2) \big], \\
    G_E^{(i)} (Q^2) &= \frac{m_N}{3(m_N + m_{\Delta_i})} \big[ \big( m_{\Delta_i} - m_N^2 -Q^2) \frac{G_1^{(i)}(Q^2)}{m_{\Delta_i}} \\
    &+ (m_{\Delta_i}^2 - m_{N}^2) G_2^{(i)} - 2Q^2 G_3^{(i)}(Q^2) \big], \\
    G_C^{(i)} (Q^2) &= \frac{2 m_N}{3(m_N + m_{\Delta_i})} \big[ 2 m_{\Delta_i} G_1^{(i)}(Q^2) + \frac{1}{2}(3 m_{\Delta_i}^2 
    + m_N^2 +Q^2)G_2^{(i)}(Q^2) \\
    &+ (m_{\Delta_i}^2 - m_{N}^2 -Q^2) G_3^{(i)}(Q^2) \big]. \\
  \end{split}
\end{equation}
After these preliminary remarks, we can proceed with the determination of these form factors for $\gamma^* N \rightarrow \Delta(1600)$ transitions within the light-cone QCD sum rules. For this purpose, we consider the following correlator function
\begin{equation}
  \label{eq:3}
\Pi_{\alpha \mu} = i \int d^4 x e^{i q x} \langle T \{ \eta_\alpha(0) j_\mu ^{el} (x) \} N(p) \rangle,
\end{equation}
where $\eta_\alpha$ is the interpolating current with the same quantum numbers of $\Delta(1232)$ and $\Delta(1600)$, and $j_\mu^{el}$ is the electromagnetic current.

Since $\Delta^+(1232)$ and $\Delta^+(1600)$ states have the same quantum numbers, the interpolating current for these states is also same and it is given by the following expression
\begin{equation}
  \label{eq:4}
  \eta_\alpha = \frac{1}{\sqrt{3}} \epsilon^{abc} \big[ 2(u^a C \gamma_\alpha d^b) u^c + (u^a C \gamma_\alpha u^b) d^c \big],
\end{equation}
where $a,b,c$ are color indices, and $C$ is the charge conjugation operator. According to the sum rules method approach, the correlation function should be calculated in two different regions. In one domain, the correlation function is saturated by the full tower of states carrying the quantum numbers of $\Delta$ baryon in the region $p{^\prime}^2 \simeq m_{\Delta_i}^2$. On the other hand, the correlation function is calculated in the deep Euclidean region where $p{^\prime}^2 << 0$ by using the operator product expansion (OPE) in terms of the nucleon distribution amplitudes with an increasing twist. The sum rules for the relevant physical quantities are obtained by matching these results of the representations of the correlation functions via the dispersion relation.

Following the mentioned prescription above and for the hadronic part of the correlation function after isolating the
contributions of the ground $\Delta(1232)$, and its first radial excitation $\Delta(1600)$ state we get
\begin{equation}
  \label{eq:5}
  \Pi_{\alpha \mu} = - \sum_{i=1}^{2} \frac{\langle 0 | \eta_\alpha(0)| \Delta_i \rangle \langle \Delta_i | j_\mu^{e l}|N \rangle}{m_{\Delta_i}^2 - p^{\prime^2}} + \int_{s_0}^\infty ds \frac{\Pi_{\alpha \mu}^{\text{had}}(s)}{s - p^{\prime^2}},
\end{equation}
where $i$ corresponds to the ground and first excited states. Parameterizing the matrix element
\begin{equation}
  \label{eq:6}
  \langle 0 | \eta_\alpha | \Delta_i(p^\prime) \rangle = \lambda_i u_\alpha (p^\prime),
\end{equation}
where $u_\alpha(p^\prime)$ is the Rarita-Schwinger spinor and $p^\prime = p - q.$ Performing summation over the spins of Rarita-Schwinger spinors with the help of the formula
\begin{equation}
  \label{eq:7}
  \sum_s u_\alpha^{(s)}(p^\prime) \bar{u}_\beta^{(s)} (p^\prime) =  - (\slashed{p}^\prime + m_{\Delta_i}) \big\{ g_{\alpha \beta} - \frac{1}{3}\gamma_\alpha \gamma_\beta - \frac{2p_\alpha^{\prime} p_\beta^{\prime} }{3 m_{\Delta_i}^2} + \frac{p_\alpha^{\prime} \gamma_\beta - p_\beta^{\prime} \gamma_\alpha}{3 m_{\Delta_i}} \big\}
\end{equation}
we get the following result for the correlation function
\begin{equation}
  \label{eq:8}
  \begin{split}
    \Pi_{\alpha \mu} =& - \sum_i \frac{\lambda_i}{m_{\Delta_i}^2 - p^{\prime 2}}(\slashed{p}^\prime + m_{\Delta_i}) \big\{ g_{\alpha \beta} - \frac{1}{3} \gamma_\alpha \gamma_\beta - \frac{2 p_\alpha^\prime p_\beta^\prime}{3m_{\Delta_i}^2} + \frac{p_\alpha^\prime \gamma_\beta - p_\beta^\prime \gamma_\alpha}{3 m_{\Delta_i}} \big\} \\
    & \big\{ G_1^{i} ( - q_\beta \gamma_\mu  + g_{\beta \mu} \slashed{q}  ) +
    G_2^{i} (-q_\beta \mathcal{P}_\mu + g_{\beta \mu} q \mathcal{P}   ) +
    G_3^{i} (q_\beta q_\mu - g_{\beta \mu} q^2 ) \big\} \gamma_5 u_N(p).
\end{split}
\end{equation}

At this point, we would like to make the following remark. In general, the interpolating current, $\eta_\alpha$, interacts not only with spin-$3/2$ states, but also with the spin-$1/2$ ones. For the generic spin-$1/2$ states, the matrix element of the $\eta_\mu$ current between the vacuum and spin 1/2 state is determined as
\begin{equation}
  \label{eq:9}
  \langle 0 | \eta_\alpha | \frac{1}{2} (p^\prime) \rangle = (m \gamma_\alpha - 4 p^\prime_\alpha) u(p^\prime). 
\end{equation}

In other words, the terms with $\sim \gamma_\alpha$ and $p_\alpha^{\prime}$ contain the contributions of the spin-$1/2$ states. Comparing 
Eqs.~\eqref{eq:8} and \eqref{eq:9}, it follows that only the terms with $\sim g_{\alpha \beta}$ contains the information of purely spin-$3/2$ states. Hence, for our problem the terms containing spin-$1/2$ contributions should be removed. Retaining the contributions of spin-$3/2$ states only, we get
\begin{equation}
  \label{eq:10}
  \begin{split}
    \Pi_{\alpha \mu} =& -  \frac{\lambda_0}{m_0^2 - {p^\prime}^2} (\slashed{p}^\prime + m_0) \big[ G_1 ( - q_\alpha \gamma_\mu + g_{\alpha \mu} \slashed{q})  \\ &+
    G_2 [ -q_\alpha (p^\prime + q/2)_\mu + q \cdot (p^\prime + q/2) g_{\alpha \mu} ]+
    G_3 [q_\alpha q_\mu - q^2 g_{\alpha \mu}] \big] \gamma_5 u_N(p)  \\& -
    \frac{\lambda_{1}}{m_1^2 - {p^\prime}^2} (\slashed{p}^\prime + m_1) \big[ \widetilde{G_1} (-q_\alpha \gamma_\mu + g_{\alpha \mu} \slashed{q}) \\ &+
    \widetilde{G_2} [ -q_\alpha (p^\prime + q/2)_\mu + q \cdot (p^\prime + q/2) g_{\alpha \mu} ] +
    \widetilde{G_3} [q_\alpha q_\mu - q^2 g_{\alpha \mu}] \big] \gamma_5 u_N(p) .
  \end{split}
\end{equation}
in which $\lambda_0,~m_0~(\lambda_1,~m_1)$ are the residue and mass of the ground state, $\Delta(1232)$, $\Delta(1600)$ baryons and $G_i(\widetilde{G_i})$ are the form factors for $\gamma^* N \rightarrow \Delta(1232)$ and $\gamma^* N \rightarrow \Delta(1600)$ transitions, respectively. For simplicity, the mass of the $\Delta(1600)$ state we will be denote as $m_1$ from now on.

From Eq.~\eqref{eq:10}, it follows that, for the description $\gamma^* N \rightarrow \Delta(1600)$ transition we have six form factors which should be determined. To determine the six form factors, we need six structures. It should be noted that all structures are not independent. To obtain the independent structures, the ordering procedure of the Dirac matrices is implemented. In this work, we choose the following order of Dirac matrices $\gamma_\alpha \slashed{p}^\prime \slashed{q} \gamma_\mu \gamma_5$. Taking into account this remark, the correlation function can be decomposed in terms of the following independent invariant functions as follows (see Eq.\eqref{eq:8}):
\begin{equation}
  \label{eq:11}
  \begin{split}
    \Pi_{\alpha \mu} =& \Pi_1 \slashed{p}^\prime \slashed{q} \gamma_5 g_{\alpha \mu} +
    \Pi_2 \slashed{q}  \gamma_5 g_{\alpha \mu} +
    \Pi_3 \slashed{p}^\prime \gamma_5 p^\prime_\mu q_\alpha +
    \Pi_4 \gamma_5 p^\prime_\mu q_\alpha +
    \Pi_5 \slashed{p}^\prime \gamma_5 q_\alpha q_\mu \\
    &+ 
    \Pi_6 \gamma_5 q_\alpha q_\mu +
    ~\text{other structures}
  \end{split}
\end{equation}

From Eqs.~\eqref{eq:9} and \eqref{eq:10}, the following six structures are found to determine the six form factors
\begin{equation}
  \label{eq:12}
  \begin{split}
   \Pi_1 &=  -  \frac{ \lambda_0 G_1}{m_0^2-p^{\prime 2}} - \frac{\lambda_1 \widetilde{G_1}}{m_1^2-p^{\prime 2}},  \\
   \Pi_2 &=  -\frac{\lambda_0 m_0 G_1}{m_0^2-p^{\prime 2}} - \frac{\lambda_1 m_1 \widetilde{G_1}}{m_1^2-p^{\prime 2}},  \\
   \Pi_3 &=  \frac{\lambda_0 G_2}{m_0^2-p^{\prime 2}} + \frac{\lambda_1 \widetilde{G_2}}{m_1^2-p^{\prime 2}},  \\
   \Pi_4 &=  \frac{\lambda_0 m_0 G_2}{m_0^2-p^{\prime 2}} + \frac{\lambda_1 m_1 \widetilde{G_2}}{m_1^2-p^{\prime 2}},  \\
   \Pi_5 &=  \frac{\lambda_0}{m_0^2-p^{\prime 2}} [\frac{G_2}{2} - G_3] + \frac{\lambda_1}{m_1^2-p^{\prime 2}} [\frac{\widetilde{G_2}}{2} - \widetilde{G_3}],  \\
   \Pi_6 &=  \frac{\lambda_0 m_0}{m_0^2-p^{\prime 2}} [\frac{G_2}{2} - G_3] + \frac{\lambda_1 m_1}{m_1^2-p^{\prime 2}} [\frac{\widetilde{G_2}}{2} - \widetilde{G_3}].
 \end{split}
\end{equation}

Solving these equations for the form factors we get
\begin{equation}
  \label{eq:13}
  \begin{split}
    - m_0 \Pi_1 + \Pi_2 &= -\frac{ \lambda_1 \widetilde{G_1} }{m_1^2 - p^{\prime 2}}(m_1 - m_0), \\
    - m_0 \Pi_3 + \Pi_4 &= \frac{\lambda_1 \widetilde{G_2}}{m_1^2 - p^{\prime 2}}(m_1 - m_0), \\
    - m_0 \Pi_5 + \Pi_6 &= \frac{\lambda_1}{m_1^2 - p^{\prime 2}}(m_1 - m_0) [\frac{\widetilde{G_2}}{2} - \widetilde{G_3}]. 
  \end{split}
\end{equation}

From Eq.\eqref{eq:13}, it follows that to obtain the sum rules for the $\gamma^* N \rightarrow \Delta(1600)$ transition form factors, we need to know the invariant functions $\Pi_i$. According to the sum rules methodology, the invariant functions $\Pi_i(i = 1 \div 6)$  are calculated at deep Euclidean domain with virtuality $p^{\prime^2} = (p - q)^2 << 0$ in terms of the nucleon distribution amplitudes (DA's). The nucleon DA's are the main non-perturbative ingredients of LCSR and they are calculated up to twist-6 in~\cite{Braun:2000kw,Wein:2015oqa,Bali:2015ykx,Anikin:2013aka,Braun:2006hz}. For completeness, definition of the nucleon's DA's and their expressions are presented in Appendix A.

Using the expressions of the nucleon DA's and applying the quark-hadron duality ansatz, the invariant functions, $\Pi_i$, can be written in the following form
\begin{equation}
  \label{eq:14}
  \Pi_i\big( p{^\prime}^2,q^2) = \sum_{n=1}^{3} \int_0^1 dx~\frac{\rho_i^{(n)} \big(x,q^2,p{^\prime}^2) \big)}{\big( (q-px)^2 \big)^n}
\end{equation}
Matching the representations of the correlation functions and performing Borel transformations with respect to the variable $-p{^\prime}^2 = -(p-q)^2$ in order to suppress the contributions of higher states and continuum, the corresponding sum rules for the form-factors
$\widetilde{G_1}(Q^2)$, $\widetilde{G_2}(Q^2)$ and $\frac{\widetilde{G_2}(Q^2)}{2} - \widetilde{G_3}(Q^2)$ can be obtained as
\begin{equation}
  \label{eq:15}
  \begin{split}
    - \lambda_1 \widetilde{G_1}(Q^2)(m_1 - m_0) e^{-m_1^{^2}/M^2} &= - m_0 I_1(M^2,Q^2,s_0) + I_2(Q^2,M^2,s_0), \\
    \lambda_1 \widetilde{G_2}(Q^2)(m_1 - m_0) e^{-m_1^{^2}/M^2} &= - m_0 I_3(M^2,Q^2,s_0) + I_4(Q^2,M^2,s_0), \\
    \lambda_1 (\frac{\widetilde{G_2}}{2} - \widetilde{G_3})(m_1 - m_0) e^{-m_1^{^2}/M^2} &= - m_0 I_5(M^2,Q^2,s_0) + I_6(Q^2,M^2,s_0). 
  \end{split}
\end{equation}
The functions $I_i(M^2,Q^2,s_0)$ can be written in the form of a master formula (see \cite{Gubernari:2018wyi} and \cite{Aliev:2019ojc})
\begin{equation}
  \label{eq:20}
  \begin{split}
    I_i =& \sum_{n=1}^\infty  \int_{x_0}^{1} dx e^{-s/M^2} \frac{1}{(n-1)!} \frac{ (-1)^n \rho_i^{(n)}}{x^n (M^2)^{n-1}}  \\
    &+ e^{-s_0/M^2}\bigg[ \frac{(-1)^{n-1}}{(n-1)!}  \sum_{j=1}^{n-1} \frac{1}{(M^2)^{n-j-1}}\frac{1}{s^\prime}\big(\frac{d}{dx} \frac{1}{s^\prime}\big)^{j-1} \frac{\rho_i^{(n)}}{x^n}  \bigg]_{ |_{x = x_0}}  \end{split}
\end{equation}
where $\bar{x} = 1-x$, $s^\prime = \frac{ds}{dx}$, $s = \frac{m_{N}^2 \bar{x} x + Q^2 \bar{x}}{x}$, and $x_0$ is the solution of $s_0 = s$ equation.
The explicit forms of the functions, $\rho_i^{(n)}$ are presented in the Appendix B. 
From Eq.\eqref{eq:15}, we see that to determine the $\gamma^* N \rightarrow \Delta(1600)$ transition form factors, the residue of $\Delta(1600)$ is also needed. This value within QCD sum rules is already calculated in~\cite{Aliev:2016jnp} and obtained as $\lambda_{1} = (0.057 \pm 0.016)~GeV^3$.

At the end of this section, we present the ratios $R_{EM}$~\cite{Jones:1972ky} and $R_{SM}$~\cite{Braun:2000kw} that are more suitable for the experimental point of view
\begin{equation}
  \label{eq:21}
  \begin{split}
    R_{EM} &= -\frac{\tilde{G}_E(Q^2)}{\tilde{G}_M(Q^2)}, \\
    R_{SM} &= - \sqrt{Q^2 + \frac{m_{1}^2 - m_N^2 - Q^2)^2}{4 m_{1}^2}} \frac{1}{2m_{1}} \frac{\tilde{G}_C(Q^2)}{\tilde{G}_M(Q^2)}.\\
   \end{split}
\end{equation}
It should be noted that these ratios are identically zero in SU(6) symmetric constituent quark model. The nonzero values are the indications of the deformation of one or both hadrons.

\section{Numerical Analysis}
\label{sec:3}
This section is devoted to the numerical analysis of the multipole form factors as well as $R_{EM}$ and $R_{SM}$ ratios. The main non-perturbative input parameters of LCSR are the DA's. In numerical calculations, for nucleon DA's we will use the results of~\cite{Wein:2015oqa,Bali:2015ykx,Anikin:2013aka,Braun:2006hz}, where the general expressions of DA's in terms of the orthogonal polynomials are obtained for octet baryons. The first few polynomials are obtained in~\cite{Anikin:2013aka}. The parameters entering in expressions of DA's are determined in~\cite{Anikin:2013aka}.

In addition to these input parameters, the sum rules contain two auxiliary parameters; the Borel parameter $M^2$ and continuum threshold $s_0$. The physically measurable quantities should be independent on these auxiliary parameters. Therefore, the working regions of $M^2$ and $s_0$ should be determined in such a way that the physically measurable quantity should exhibit good stability to the variation of these parameters. The upper and lower bounds of the Borel parameter $M^2$ are determined by imposing the following two conditions.
\begin{itemize}
\item The reasonable suppression of the integral over the higher states contributions estimated in accordance of the hadron-quark duality ansatz.
  \item Contributions of the higher twist terms should be smaller than the contributions of the leading twist term.
\end{itemize}
Besides, the values of continuum threshold $s_0$ is determined from the condition that the sum rules should reproduce the mass of $\Delta(1600)$ state with $10\%$ accuracy. These conditions lead to the following working regions of $M^2$ and $s_0$ : $2.0~\rm{GeV^2} \leq M^2 \leq 4.0~\rm{GeV^2}$, $s_0 = (5.5 \pm 0.5)~\rm{GeV^2}.$

Having specified all the input parameters and determined the working region of $M^2$ and $s_0$, we are ready to perform the numerical calculations.

\begin{figure}[h!]
  \centering
  \includegraphics[scale=0.60]{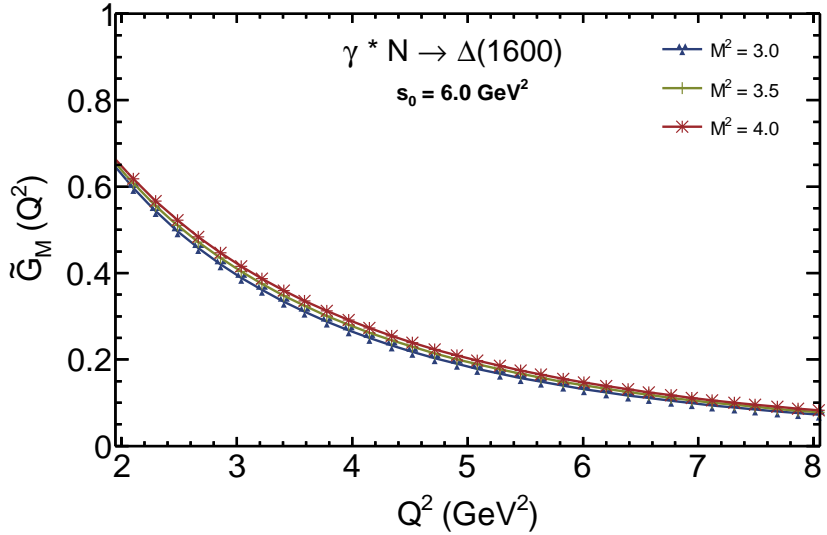}
  \caption{The dependency of the $\widetilde{G_M}(Q^2)$ on $Q^2$ at a fixed values of $s_0$ and $M^2$.}
  \label{fig:1}
\end{figure}

\begin{figure}[hbt!]
  \centering
  \includegraphics[scale=0.60]{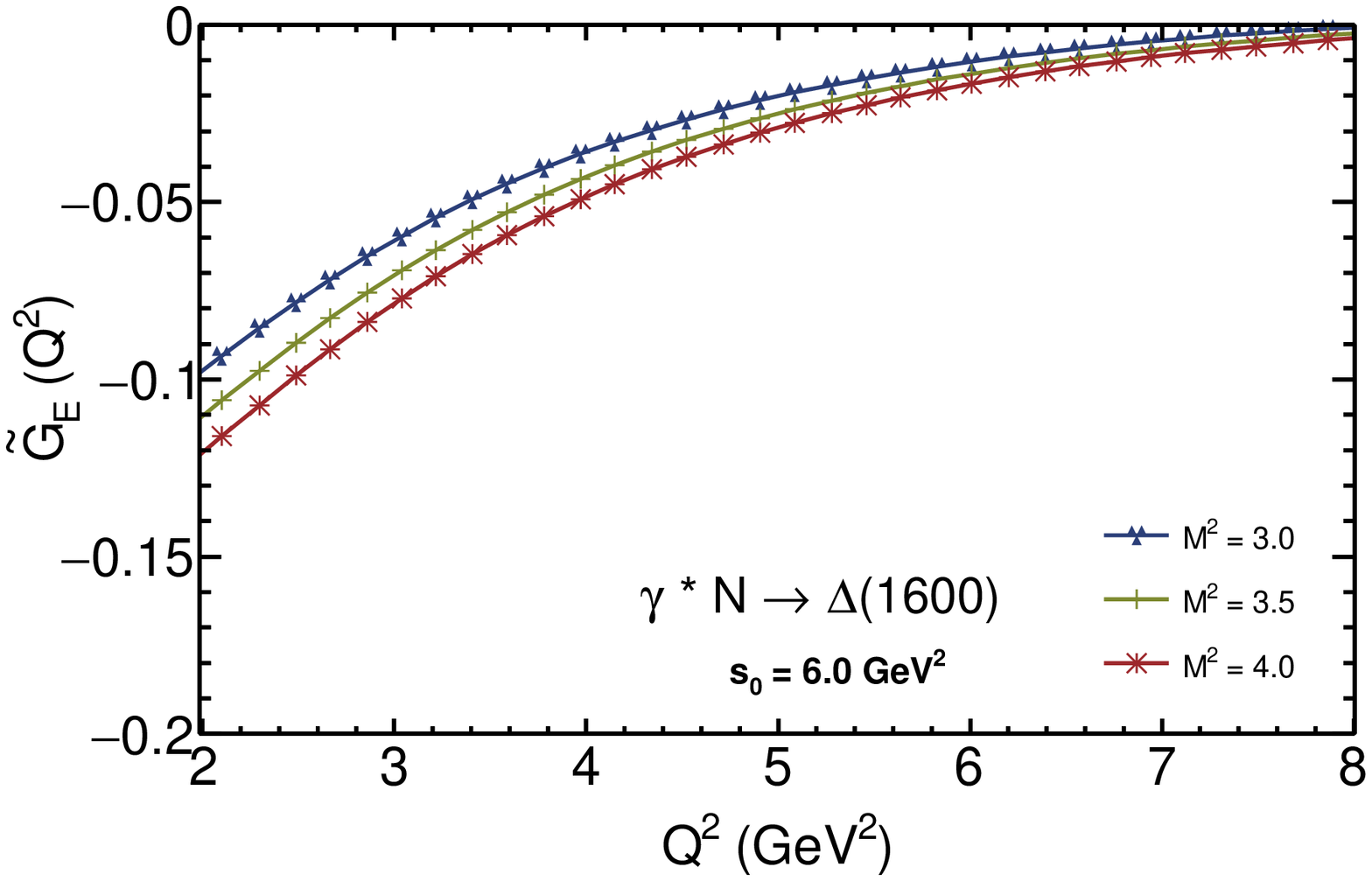}
  \caption{The same as in Fig.~\ref{fig:1}, but for $\widetilde{G_E}(Q^2)$ form factor.}
  \label{fig:2}
\end{figure}

\begin{figure}[hbt!]
  \centering
  \includegraphics[scale=0.60]{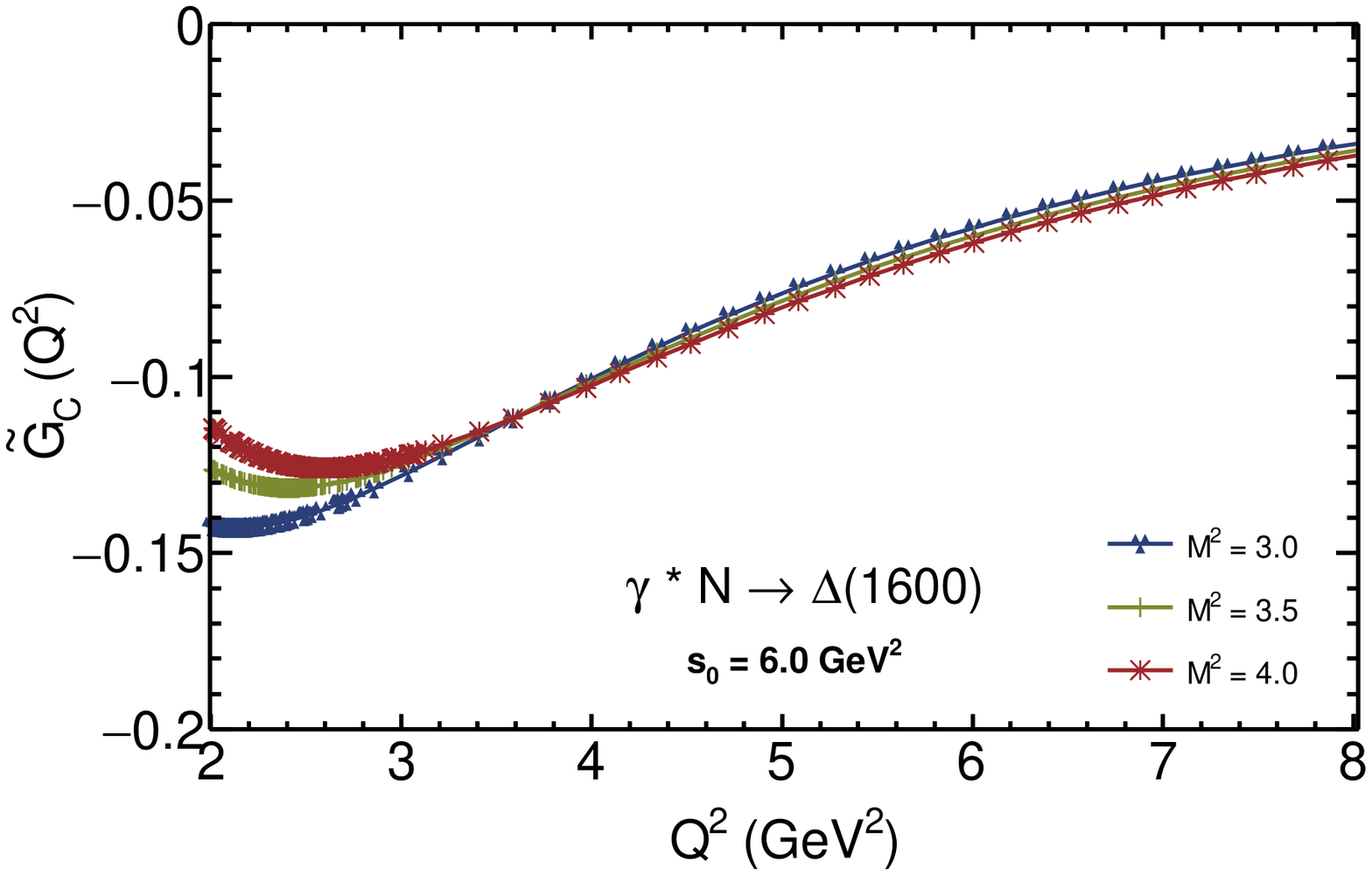}
  \caption{The same as in Fig.~\ref{fig:1}, but for $\widetilde{G_C}(Q^2)$ form factor.}
  \label{fig:3}
\end{figure}

In Figures, \ref{fig:1},~\ref{fig:2}, and \ref{fig:3}, we present the dependencies of $\tilde{G}_{M}(Q^2)$, $\tilde{G}_E(Q^2)$ and $\tilde{G}_C(Q^2)$ on $Q^2$ at fixed $s_0$ and for various $M^2$ values. Here, we would like to note that since LCSR predictions on the form factors are reliable only in the $Q^2 > 1~\rm{GeV^2}$ region,  we present the results only for this domain. From these figures, it follows that all three form factors decrease with increasing $Q^2$ and saturates for high $Q^2$ values.

By comparing the form factors of $\gamma^* N \rightarrow \Delta(1232)$ obtained in~\cite{Braun:2005be} and $\gamma^* N \rightarrow \Delta(1600)$ transitions, we infer the following results :
\begin{itemize}
\item The electric quadrupole form factor is small in magnitude compared to the form factors, $\tilde{G}_M(Q^2)$ and $\tilde{G}_C(Q^2)$ in both transitions. 
  \item For the region, $Q^2 > 1~\rm{GeV^2}$, the transition form factors, $\tilde{G}_M(Q^2)$ for $\gamma^* N \rightarrow \Delta(1600)$ are larger than the ones for $\gamma^* N \rightarrow \Delta(1232)$. This result indicated that the $\gamma^* N \rightarrow \Delta^{+}(1600)$ transition is more localized in configuration space. 
\end{itemize}

Moreover, we also compared our results on the considered form factors with the predictions obtained by quark-diquark approximation to the Poincare-covariant three-body bound state problem in relativistic quantum field theory~\cite{Lu:2019bjs} and found out that our predictions on the form factors at the considered regions of $Q^2$ is considerably larger in magnitude than the one predicted in~\cite{Lu:2019bjs}.

\begin{figure}[hbt!]
  \centering
  \includegraphics[scale=0.60]{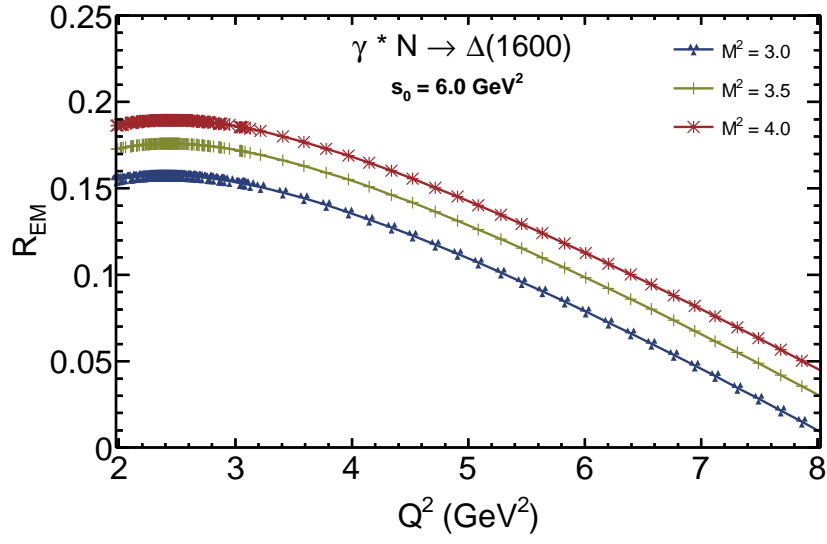}
  \caption{The dependency of $R_{EM}$ on $Q^2$ at the fixed values of $s_0$ and $M^2$.}
  \label{fig:4}
\end{figure}

\begin{figure}[hbt!]
  \centering
  \includegraphics[scale=0.60]{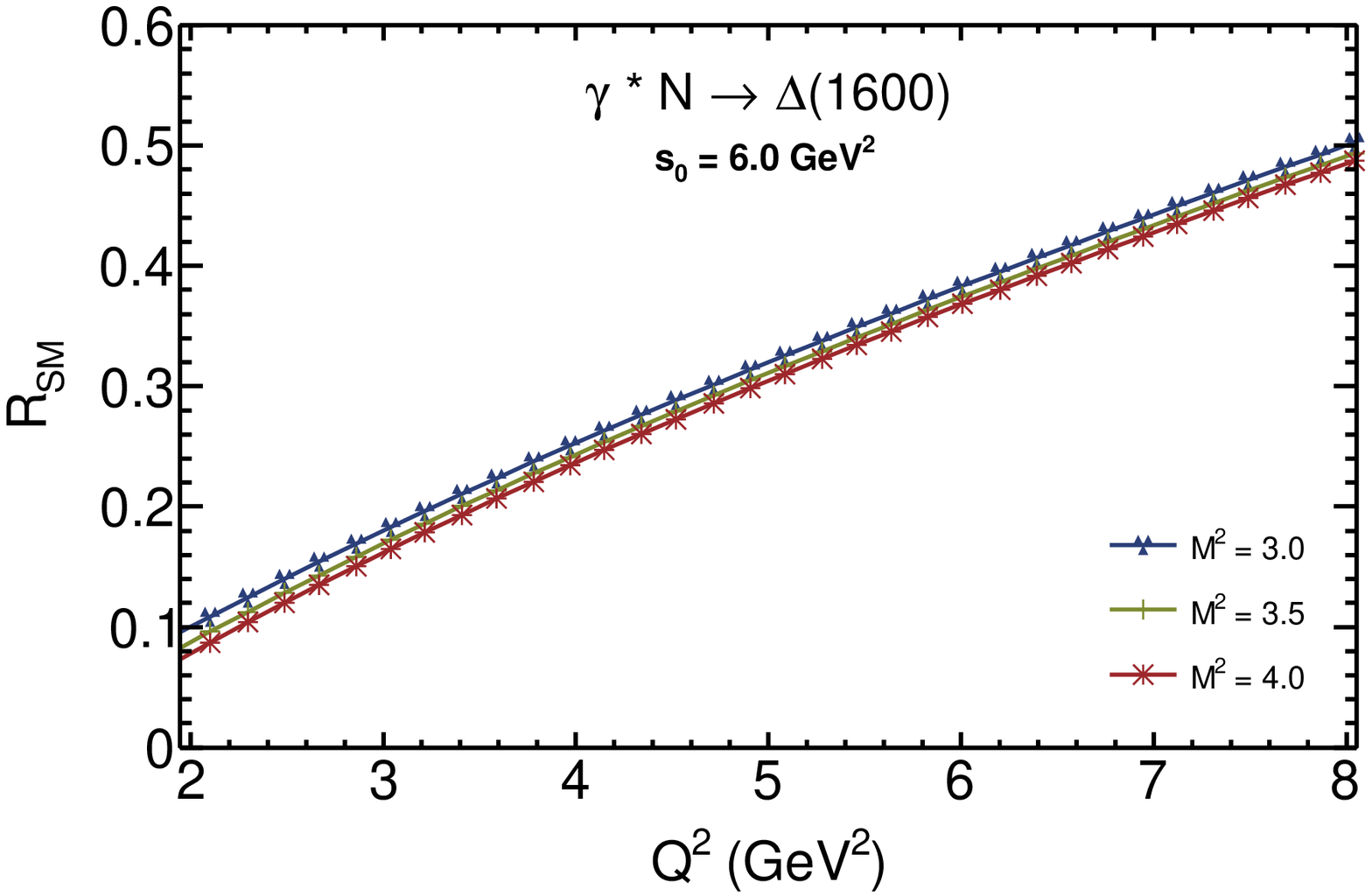}
  \caption{The same as in Fig.~\ref{fig:4}, but for $R_{SM}$.}
  \label{fig:5}
\end{figure}

Furthermore in Figures~\ref{fig:4} and \ref{fig:5}, we present the $Q^2$ dependence of  $R_{EM}(Q^2) $ and $R_{SM} $  at fixed values $M^2$ and $s_0$ considering their working regions. From these figures, we observe that while $R_{SM}(Q^2)$ is negative, $R_{EM}(Q^2)$ is positive at all values of $Q^2$.

Comparing our predictions on $R_{SM}$ with the results obtained in~\cite{Braun:2005be}, we observed similar qualitative behaviour  considered in both works. However, behavior of $R_{EM}$ in our case is remarkably different than in $\gamma^* N \rightarrow \Delta(1232)$ transition, i.e. magnitude $R_{EM}$ is larger than the one in $\gamma^* N \rightarrow \Delta(1232)$ transition case. This observation highlights the sensitivity of the electric quadrupole form factor to the degree of deformation of the $\Delta$ baryon. Finally, we compare our predictions on $R_{SM}$ and $R_{EM}$ with the results obtained within light-front relativistic quark model~\cite{Aznauryan:2015zta}. Comparing our results on $R_{EM}$ presented in~Fig.~\ref{fig:4} and the results of~\cite{Aznauryan:2015zta} we observed that our result on $R_{EM}$ is larger than the one predicted in~\cite{Aznauryan:2015zta}.  Besides, comparing our result on $R_{SM}$, we deduce that the behaviour of $R_{SM}$ is similar to the results of~\cite{Aznauryan:2015zta}. For example, in our case, when $Q^2$ varies between $2$ and $8~\rm{GeV^2}$ region, the $R_{SM}$ varies between $(0.1 ~\text{and } 0.5)$, however it changes between ($0.1~\text{and } 0.3 $) in~\cite{Aznauryan:2015zta}.

Our final note is that the obtained results will shed light to the understanding the inner structures of the resonance $\Delta(1600)$, and can be checked in ongoing and planning experiments.

\section{Conclusion}
In this article, we studied the LCSR to evaluate the magnetic dipole $\tilde{G}_M(Q^2)$ electric quadrupole $\tilde{G}_E(Q^2)$ and Coulomb quadrupole $\tilde{G}^*(Q^2)$ form factors as well as the ratios $R_{EM} = -\frac{\tilde{G}_E}{\tilde{G}_M}$ and $R_{SM} = - \frac{1}{4 m_1^2} \sqrt{4 m_1^2 Q^2 + (m_1^2 - Q^2 -m_N^2)^2} \frac{\tilde{G}_c(Q^2)}{\tilde{G}_M(Q^2)}$ on $Q^2$ when $Q^2$ varies in the region $1~\rm{GeV}^2 \leq Q^2 \leq ~10~{GeV^2}$. This domain may be covered in the incoming CLAS-12 at the Jefferson Lab. Appearance of experimental information would be very useful to establish the nature of $\Delta^+(1600)$ resonance by assuming it as radial excitation of $\Delta(1232)$ in the $\gamma^* N \rightarrow \Delta(1600)$ transition.
We also compared our predictions on the form factors $\tilde{G}_M(Q^2)$, $\tilde{G_E}(Q^2)$, and $\tilde{G_C}(Q^2)$, as well as $R_{EM}$ and $R_{SM}$ with results of results of other theoretical approaches. 

\section*{Acknowledgment}
One of the authors, T.Barakat extends his appreciation to the Deanship of Scientific Research at King Saud University for funding his work through Research Grant No: RG-1440-090.


\clearpage


\bibliographystyle{apsrev4-1}
\bibliography{multipole}


\section*{Appendix A: Nucleon Distribution Amplitudes}
\label{sec:appendixA}
\setcounter{equation}{0}

For completeness, in this Appendix, we present expressions of distribution amplitudes $V_i, A_i, T_i, S_i,$ and $P_i$ for nucleon.

\begin{align*}
V_{1/6}&= \frac{1}{2}\biggl(\Phi_{+,3/6}{\scriptstyle(x_1,x_2,x_3)} + \Phi_{-,3/6}{\scriptstyle(x_1,x_2,x_3)} \biggr) + \frac{1 }{2} \biggl( \Phi_{+,3/6}{\scriptstyle(x_2,x_1,x_3)} + \Phi_{-,3/6}{\scriptstyle(x_2,x_1,x_3)}  \biggr)  \ , \nonumber \\
A_{1/6} &=   - \frac{1}{2}\biggl(\Phi_{+,3/6}{\scriptstyle(x_1,x_2,x_3)} + \Phi_{-,3/6}{\scriptstyle(x_1,x_2,x_3)} \biggr) + \frac{1 }{2} \biggl( \Phi_{+,3/6}{\scriptstyle(x_2,x_1,x_3)} + \Phi_{-,3/6}{\scriptstyle(x_2,x_1,x_3)}  \biggr) \ , \nonumber \\
T_{1/6} &=  \Pi_{3/6}{\scriptstyle(x_1,x_3,x_2)} \ , 
\end{align*}
where the DAs on the l.h.s.\ are functions of $(x_1,x_2,x_3)$. The twist $4$ and twist $5$ amplitudes read%

\begin{align*}
S _{1/2} &= \frac{1 }{24}\biggl( \Xi _{+,4/5}{\scriptstyle(x_1,x_2,x_3)} +  \Xi _{-,4/5}{\scriptstyle(x_1,x_2,x_3)}  \biggr) - \frac{1 }{24}\biggl(  \Xi _{+,4/5}{\scriptstyle(x_2,x_1,x_3)}  +  \Xi _{-,4/5}{\scriptstyle(x_2,x_1,x_3)}   \biggr) \nonumber \\*
&\quad + \frac{1}{4} \biggl( \Pi _{4/5}{\scriptstyle(x_2,x_3,x_1)}  -   \Pi _{4/5}{\scriptstyle(x_1,x_3,x_2)}  \biggr) \ , \nonumber \\
P _{1/2} &= \frac{1 }{24}\biggl( \Xi _{+,4/5}{\scriptstyle(x_1,x_2,x_3)}  +  \Xi _{-,4/5}{\scriptstyle(x_1,x_2,x_3)}  \biggr) - \frac{1 }{24}\biggl(  \Xi _{+,4/5}{\scriptstyle(x_2,x_1,x_3)}  +  \Xi _{-,4/5}{\scriptstyle(x_2,x_1,x_3)}   \biggr) \nonumber \\*
&\quad - \frac{1}{4} \biggl( \Pi _{4/5}{\scriptstyle(x_2,x_3,x_1)}  -   \Pi _{4/5}{\scriptstyle(x_1,x_3,x_2)}  \biggr) \ , \nonumber \\
V _{2/5} &= \frac{1}{4}\biggl( \Phi _{+,4/5}{\scriptstyle(x_1,x_2,x_3)}  +  \Phi _{-,4/5}{\scriptstyle(x_1,x_2,x_3)}  \biggr) + \frac{1 }{4} \biggl(  \Phi _{+,4/5}{\scriptstyle(x_2,x_1,x_3)}  +  \Phi _{-,4/5}{\scriptstyle(x_2,x_1,x_3)}   \biggr) \ , \nonumber \\
A _{2/5} &= - \frac{1}{4}\biggl( \Phi _{+,4/5}{\scriptstyle(x_1,x_2,x_3)}  +  \Phi _{-,4/5}{\scriptstyle(x_1,x_2,x_3)}  \biggr) + \frac{1 }{4} \biggl(  \Phi _{+,4/5}{\scriptstyle(x_2,x_1,x_3)}  +  \Phi _{-,4/5}{\scriptstyle(x_2,x_1,x_3)}   \biggr) \ , \nonumber \\
V _{3/4} &= \frac{1}{4}\biggl( \Phi _{+,4/5}{\scriptstyle(x_3,x_1,x_2)}  -  \Phi _{-,4/5}{\scriptstyle(x_3,x_1,x_2)}  \biggr) + \frac{1 }{4} \biggl(  \Phi _{+,4/5}{\scriptstyle(x_3,x_2,x_1)}  -  \Phi _{-,4/5}{\scriptstyle(x_3,x_2,x_1)}   \biggr) \ , \nonumber \\
A _{3/4} &= - \frac{1}{4}\biggl( \Phi _{+,4/5}{\scriptstyle(x_3,x_1,x_2)}  -  \Phi _{-,4/5}{\scriptstyle(x_3,x_1,x_2)}  \biggr) + \frac{1 }{4} \biggl(  \Phi _{+,4/5}{\scriptstyle(x_3,x_2,x_1)}  -  \Phi _{-,4/5}{\scriptstyle(x_3,x_2,x_1)}   \biggr) \ , \nonumber \\
T _{2/5} &=  \frac{\Upsilon _{4/5}{\scriptstyle(x_3,x_2,x_1)}}{6 } \ , \nonumber \\
T _{3/4} &= \frac{1 }{24}\biggl( \Xi _{+,4/5}{\scriptstyle(x_1,x_2,x_3)}  +  \Xi _{-,4/5}{\scriptstyle(x_1,x_2,x_3)}  \biggr) + \frac{1 }{24}\biggl(  \Xi _{+,4/5}{\scriptstyle(x_2,x_1,x_3)}  +  \Xi _{-,4/5}{\scriptstyle(x_2,x_1,x_3)}   \biggr) \nonumber \\*
&\quad + \frac{1}{4} \biggl( \Pi _{4/5}{\scriptstyle(x_2,x_3,x_1)}  + 1  \Pi _{4/5}{\scriptstyle(x_1,x_3,x_2)}  \biggr) \ , \nonumber \\
T _{7/8} &= -\frac{1 }{24}\biggl( \Xi _{+,4/5}{\scriptstyle(x_1,x_2,x_3)}  +  \Xi _{-,4/5}{\scriptstyle(x_1,x_2,x_3)}  \biggr) - \frac{1 }{24}\biggl(  \Xi _{+,4/5}{\scriptstyle(x_2,x_1,x_3)}  +  \Xi _{-,4/5}{\scriptstyle(x_2,x_1,x_3)}   \biggr) \nonumber \\*
&\quad + \frac{1}{4} \biggl( \Pi _{4/5}{\scriptstyle(x_2,x_3,x_1)}  +   \Pi _{4/5}{\scriptstyle(x_1,x_3,x_2)}  \biggr) \ .
\end{align*}%

The explicit expressions of the functions $\Phi_{\pm,3/6}$, $\Xi _{\pm,4/5}$, $\Phi_{\pm,4/5}$, $\Upsilon_{4/5}$, and $\Pi _{4/5}$ can be found in~\cite{Wein:2015oqa,Bali:2015ykx,Anikin:2013aka}.

\section*{Appendix B: Correlation Functions}
\label{sec:appendixB}
\setcounter{equation}{0}
In this Appendix we present the explicit expressions of the functions
$\rho_i^n$ entering to the sum rules for the form factors $\widetilde{G}_1(Q^2)$,
$\widetilde{G}_2(Q^2)$ and $ \frac{\widetilde{G}_2(Q^2)}{2} - \widetilde{G}_3(Q^2)$ for the
$\gamma^\ast N \to \Delta(1600)$ transition. 

\section*{Functions $\rho_i^{(n)}$ for the form factor $\widetilde{G_1}$}

\begin{equation}
  \label{eq:100}
  \begin{split}
\rho_{1}^{(3)} (x)  &= 0 \\
\rho_{2}^{(3)} (x) &=  \frac{8 (1-x)}{x} e_{q_2}   m_{N}^2 m_{q_2} (x^2  m_{N}^2+Q^{2}) \, \widetilde{\!\widetilde{B}}_6     \\
\rho_{1}^{(2)} (x) &=  -4 e_{q_3}   m_{N} ( m_{N} \widehat{\!\widehat{B}}_6-2 m_{q_3} \widehat{B}_4) + 8 e_{q_2}   m_{N} m_{q_2} \widetilde{B}_2 \\
&-8 e_{q_2}   m_{N}^2 \int_0^{\bar{x}} dx_1 
 ( {T_1}^{M}-{A_1}^{M}) ( x_1, x, 1-x_1 - x) \\
 &+8 e_{q_3}   m_{N}^2 \int_0^{\bar{x}} dx_1 
{T_1}^{M} ( x_1, 1-x_1 - x, x) \\
\rho_{2}^{(2)} (x)  &=  - \frac{4  m_{N}}{x} \Big\{
 -e_{q_1}  \Big[(x-1) (x^2  m_{N}^2+Q^{2}) \check{C}_2 + 2 x (x+1) \check{D}_2 \Big] \\
 &+e_{q_2}  \Big[ x^3  m_{N}^2 \widetilde{B}_2+x^3  m_{N}^2 \widetilde{B}_4+(x-1) (x^2  m_{N}^2+Q^{2}) \widetilde{D}_2-(x-1) (x^2  m_{N}^2+Q^{2}) \widetilde{C}_2 \\
 &-x^2  m_{N}^2 \widetilde{B}_2-x^2  m_{N}^2 \widetilde{B}_4-2 x^2  m_{N} m_{q_2} \widetilde{H}_1+2 x^2  m_{N} m_{q_2} \widetilde{B}_5+2 x^2  m_{N} m_{q_2} \widetilde{B}_7 \\
 &+2 (x-1) x  m_{N}^2 \, \widetilde{\!\widetilde{B}}_8+2 x  m_{N} m_{q_2} \widetilde{H}_1-2 x  m_{N} m_{q_2} \widetilde{B}_5-2 x  m_{N} m_{q_2} \widetilde{B}_7-x  m_{N} m_{q_2} \, \widetilde{\!\widetilde{B}}_6 \\
 &+x Q^{2} \widetilde{B}_2+x Q^{2} \widetilde{B}_4-2  m_{N} m_{q_2} \, \widetilde{\!\widetilde{B}}_6-Q^{2} \widetilde{B}_2-Q^{2} \widetilde{B}_4 \Big]  \\
 &+x e_{q_3}   m_{N} \Big[ (x-1) \Big( m_{N} ( \, \widehat{\!\widehat{D}}_6-2 \; \widehat{\!\widehat{C}}_6) 
 +m_{q_3} (\widehat{D}_5-2 \widehat{C}_5+2 \widehat{B}_5+4 \widehat{B}_7) \Big)  + m_{q_3} \, \widehat{\!\widehat{B}}_6 \Big]
 \Big\} \\
\rho_{1}^{(1)} (x) &= -e_{q_2} \int_0^{\bar{x}} dx_1 
  (8 {B_1}-8 {D_1}) ( x_1, x, 1-x_1 - x)
+8 e_{q_3} \int_0^{\bar{x}} dx_1 
{B_1} ( x_1, 1-x_1 - x, x)  \\
\rho_{2}^{(1)} (x) &= \frac{4 m_{N}}{x} \Big[ e_{q_2}    (\widetilde{D}_2-\widetilde{C}_2+\widetilde{B}_2+\widetilde{B}_4)- e_{q_1}    (x \check{D}_2+\check{C}_2) \Big] \\
&+ 4 (x-1) e_{q_1}   m_{N} \int_0^{\bar{x}} dx_3 ({C_3}-{D_3}) ( x , 1-x - x_3,x_3) \\
    &-4e_{q_2}   (x-1)  m_{N} \int_0^{\bar{x}} dx_1  \Big[ ({D_3}-{C_3}  +2 {P_1}-2 {S_1})-8 m_{q_2} {B_1} \Big] ( x_1, x, 1-x_1 - x) \\
    &+4 e_{q_3}   (x-1)  m_{N} \int_0^{\bar{x}} dx_1 \Big[ ({D_3}-2 {C_3})-8 m_{q_3} {B_1} \Big] ( x_1, 1-x_1 - x, x)
  \end{split}
\end{equation}

\section*{Functions $\rho_i$ for the form factor $\widetilde{G_2}$}

\begin{equation}
  \label{eq:101}
  \begin{split}
    \rho_{3}^{(3)} (x)  &= 64 (x-1) x^2 e_{q_1}   m_{N}^3 \, \check{\!\check{C}}_6 +  16 x e_{q_2}   m_{N}^2 \Big[ 4 (x-1) x  m_{N}  (\, \widetilde{\!\widetilde{C}}_6-2 \, \widetilde{\!\widetilde{B}}_8)-m_{q_2} \, \widetilde{\!\widetilde{B}}_6 \Big] \\
    &+16 x e_{q_3}   m_{N}^2 \Big[ m_{q_3} \, \widehat{\!\widehat{B}}_6-2 (x-1) x  m_{N} ( \, \widehat{\!\widehat{D}}_6-2 \; \widehat{\!\widehat{C}}_6+2 \, \widehat{\!\widehat{B}}_8) \Big]    \\
    \rho_{4}^{(3)} (x) &= 32 (x-1) e_{q_2}   m_{N}^2 \Big[ (x^2  m_{N}^2+2 x Q^{2}-Q^{2}) \, \widetilde{\!\widetilde{B}}_6+ x  m_{N} m_{q_2} \, \widetilde{\!\widetilde{B}}_8 \Big] \\
    &+16 (x-1) e_{q_3}   m_{N}^2 \Big[ 2 x (x  m_{N}^2+Q^{2} ) \, \widehat{\!\widehat{B}}_6-x  m_{N} m_{q_3} (\, \widehat{\!\widehat{D}}_6+2 \; \widehat{\!\widehat{C}}_6)\Big] \\
    \rho_{3}^{(2)} (x) &= -8 (1-2 x) x e_{q_1}   m_{N} \check{C}_2 + 8 x e_{q_2}   m_{N} \Big[ (2 x-1) \widetilde{C}_2+2 (1-2 x) \widetilde{B}_4-\widetilde{D}_2-2 \widetilde{B}_2 \Big] \\
    & -8 x e_{q_3}   m_{N} \Big[ x \widehat{D}_2-2 x \widehat{C}_2+2 (x-1) \widehat{B}_4 \Big] \\
    \rho_{4}^{(2)}(x)  &= -8 (x-1) x e_{q_1}   m_{N}^2 (\check{D}_5-\check{C}_4)-8 e_{q_2}   m_{N} \Big\{(x-1) x  m_{N} \Big[ \widetilde{D}_5-\widetilde{C}_4-2 (\widetilde{H}_1+\widetilde{E}_1-\widetilde{B}_5)\Big]\\
    &+(4 x-3)  m_{N} \, \widetilde{\!\widetilde{B}}_6+2 x m_{q_2} \widetilde{B}_4+m_{q_2} (\widetilde{B}_2-\widetilde{B}_4) \Big\}+8 x e_{q_3}   m_{N} \Big[(x-1)  m_{N} (\widehat{D}_5+2 \widehat{C}_5-2 \widehat{B}_5) \\
    &+m_{q_3} (\widehat{D}_2+2 \widehat{C}_2)\Big]
    + 8 (2 x-1) e_{q_1}   m_{N}^2 \int_0^{\bar{x}} dx_3 {V_1}^{M} ( x, 1-x - x_3,x_3) \\
    & -8 e_{q_2}   m_{N}^2 \int_0^{\bar{x}} dx_1  \Big[{A_1}^{M}+(1-2 x) {V_1}^{M}+2 (x-1) {T_1}^{M}  \Big] ( x_1, x, 1-x_1 - x) \\
   & -16 x e_{q_3}   m_{N}^2 \int_0^{\bar{x}} dx_1 
   {T_1}^{M} ( x_1, 1-x_1 - x, x) \\
   \rho_{3}^{(1)}(x)  &= 0 \\
    \rho_{4}^{(1)}(x)  &= 8 (2 x-1) e_{q_1}  \int_0^{\bar{x}} dx_3
{C_1}( x, 1-x - x_3,x_3) \\
 & -8 e_{q_2} \int_0^{\bar{x}} dx_1  \Big[{D_1}-(2 x-1) {C_1} + 2 (x-1) {B_1} \Big] ( x_1, x, 1-x_1 - x) \\
 &-
    16 x e_{q_3} \int_0^{\bar{x}} dx_1  {B_1} ( x_1, 1-x_1 - x, x)      
  \end{split}
\end{equation}

\section*{Functions $\rho_i$ for the form factor $\frac{\widetilde{G_2}}{2} - \widetilde{G_3}$}

\begin{equation}
  \label{eq:102}
  \begin{split}
    \rho_{5}^{(3)} (x)  &= -64 (x-1)^2 x e_{q_1}   m_{N}^3 \, \check{\!\check{C}}_6 - 16 (x-1) e_{q_2}   m_{N}^2 \Big[ 4 (x-1) x  m_{N} ( \, \widetilde{\!\widetilde{C}}_6-2 \, \widetilde{\!\widetilde{B}}_8)-2 m_{q_2} \, \widetilde{\!\widetilde{B}}_6 \Big] \\
    &- 16 (x-1) e_{q_3}   m_{N}^2 \Big[ 2 (x-1) x  m_{N} (2 {C_6}-\widehat{\!\widehat{D}}_6- 2 \, \widehat{\!\widehat{B}}_8)+m_{q_3} \, \widehat{\!\widehat{B}}_6\Big] \\
    \rho_{6}^{(3)} (x)  &=  \frac{- 32 (x-1)^2}{x} e_{q_2}   m_{N}^2 \Big[ (x^2  m_{N}^2+2 x Q^{2}-Q^{2}) \, \widetilde{\!\widetilde{B}}_6+x  m_{N} m_{q_2} \, \widetilde{\!\widetilde{B}}_8 \Big]\\
    & -\frac{16 (x-1)^2}{x} e_{q_3}   m_{N}^2 \Big[2 x (x  m_{N}^2+Q^{2}) \, \widehat{\!\widehat{B}}_6-x  m_{N} m_{q_3} (2 {C_6}+\widehat{\!\widehat{D}}_6)\Big] \\
    \rho_{5}^{(2)} (x)  &= - 16 (x-1) x e_{q_1}   m_{N} \check{C}_2 + 16 (x-1) e_{q_2}   m_{N} (-x \widetilde{C}_2+2 x \widetilde{B}_4+\widetilde{D}_2+\widetilde{B}_2)\\
    &+ 8 (x-1) e_{q_3}   m_{N} \Big[x \widehat{D}_2-2 x \widehat{C}_2+2 (x-1) \widehat{B}_4\Big]  \\
    \rho_{6}^{(2)} (x)  &= 8 (x-1)^2 e_{q_1}   m_{N}^2 (\check{D}_5-\check{C}_4) 
    + \frac{8 (x-1)}{x} e_{q_2}   m_{N} \Big\{ (x-1) x  m_{N} \Big[ \widetilde{D}_5-\widetilde{C}_4-2 (\widetilde{H}_1+\widetilde{E}_1-\widetilde{B}_5)\Big] \\
    &+4 (x-1)  m_{N}\, \widetilde{\!\widetilde{B}}_6+2 x m_{q_2} \widetilde{B}_4 \Big\} \\
    &-8 (x-1) e_{q_3}   m_{N} \Big[(x-1)  m_{N} (\widehat{D}_5+2 \widehat{C}_5-2 \widehat{B}_5)+m_{q_3} (\widehat{D}_2+2  \widehat{C}_2)\Big] \\
    &-16 (x-1) e_{q_1}   m_{N}^2 \int_0^{\bar{x}} dx_3 {V_1}^{M} ( x, 1-x - x_3,x_3) \\
    &-16 (x-1) e_{q_2}   m_{N}^2 \int_0^{\bar{x}} dx_1  ({V_1}^{M}-{T_1}^{M})( x_1, x, 1-x_1 - x) \\
    &+16 (x-1) e_{q_3}   m_{N}^2 \int_0^{\bar{x}} dx_ {T_1}^{M}  ( x_1, 1-x_1 - x, x) \\
    \rho_{5}^{(1)} (x)  &= 0 \\
    \rho_{6}^{(1)}(x)  &= -16 (x-1) e_{q_1} \int_0^{\bar{x}} dx_3  {C_1} ( x, 1-x - x_3,x_3) \\
    &-16 (x-1) e_{q_2} \int_0^{\bar{x}} dx_1   ({C_1}-{B_1}) ( x_1, x, 1-x_1 - x) \\
    &+16 (x-1) e_{q_3} \int_0^{\bar{x}} dx_1  {B_1} ( x_1, 1-x_1 - x, x)
  \end{split}
\end{equation}

where $q_1=u$, $q_2=u$, and $q_3=d$, respectively.

In the above expressions for $\rho_2$, $\rho_4$, and $\rho_6$
the functions ${\cal F}(x_i)$ are defined in the following way:

\bea
\label{nolabel}
\check{\cal F}(x_1) \es \int_1^{x_1}\!\!dx_1^{'}\int_0^{1-
x^{'}_{1}}\!\!dx_3\,
{\cal F}(x_1^{'},1-x_1^{'}-x_3,x_3)~, \nnb \\
\check{\!\!\!\;\check{{\cal F}}}(x_1) \es
\int_1^{x_1}\!\!dx_1^{'}\int_1^{x^{'}_{1}}\!\!dx_1^{''}
\int_0^{1- x^{''}_{1}}\!\!dx_3\,
{\cal F}(x_1^{''},1-x_1^{''}-x_3,x_3)~, \nnb \\
\widetilde{\cal F}(x_2) \es \int_1^{x_2}\!\!dx_2^{'}\int_0^{1-
x^{'}_{2}}\!\!dx_1\,
{\cal F}(x_1,x_2^{'},1-x_1-x_2^{'})~, \nnb \\
\widetilde{\!\widetilde{\cal F}}(x_2) \es
\int_1^{x_2}\!\!dx_2^{'}\int_1^{x^{'}_{2}}\!\!dx_2^{''}
\int_0^{1- x^{''}_{2}}\!\!dx_1\,
{\cal F}(x_1,x_2^{''},1-x_1-x_2^{''})~, \nnb \\
\widehat{\cal F}(x_3) \es \int_1^{x_3}\!\!dx_3^{'}\int_0^{1-
x^{'}_{3}}\!\!dx_1\,
{\cal F}(x_1,1-x_1-x_3^{'},x_3^{'})~, \nnb \\
\widehat{\!\widehat{\cal F}}(x_3) \es
\int_1^{x_3}\!\!dx_3^{'}\int_1^{x^{'}_{3}}\!\!dx_3^{''}
\int_0^{1- x^{''}_{3}}\!\!dx_1\,
{\cal F}(x_1,1-x_1-x_3^{''},x_3^{''})~.\nnb
\eea
Definitions of the functions $B_i$, $C_i$, $D_i$, $E_1$ and $H_1$
that appear in the expressions for $\rho_i(x)$ are given as follows:

\bea
\label{nolabel}
B_2 \es T_1+T_2-2 T_3~, \nnb \\
B_4 \es T_1-T_2-2 T_7~, \nnb \\
B_5 \es - T_1+T_5+2 T_8~, \nnb \\
B_6 \es 2 T_1-2 T_3-2 T_4+2 T_5+2 T_7+2 T_8~, \nnb \\
B_7 \es T_7-T_8~, \nnb \\
B_8 \es  -T_1+T_2+T_5-T_6+2 T_7+2T_8~, \nnb \\
C_2 \es V_1-V_2-V_3~, \nnb \\
C_4 \es -2V_1+V_3+V_4+2V_5~, \nnb \\
C_5 \es V_4-V_3~, \nnb \\
C_6 \es -V_1+V_2+V_3+V_4+V_5-V_6~, \nnb \\
D_2 \es -A_1+A_2-A_3~, \nnb \\
D_4 \es -2A_1-A_3-A_4+2A_5~, \nnb \\
D_5 \es A_3-A_4~, \nnb \\
D_6 \es A_1-A_2+A_3+A_4-A_5+A_6~, \nnb \\
E_1 \es S_1-S_2~, \nnb \\
H_1 \es P_2-P_1~. \nnb
\eea

The expressions of the functions $V_i$, $A_i$, $T_i$, $S_i$ and $P_i$ are presented in Appendix A.



\end{document}